\documentclass[%
 reprint,
superscriptaddress,
 amsmath,amssymb,
 aps
]{revtex4-2}
\usepackage{graphicx}% Include figure files
\usepackage{dcolumn}% Align table columns on decimal point
\usepackage{bm}% bold math
\usepackage[pagewise, modulo]{lineno}% Enable numbering of text and display math
\usepackage[utf8]{inputenc}
\usepackage[T1]{fontenc}
\usepackage{mathptmx}
\usepackage{pgfplots}
\usepackage{tikz}
\usetikzlibrary{shapes,arrows.meta,patterns,calc}
\tikzset{every picture/.style={/utils/exec={\sffamily}}}
\usepackage{helvet}
\usepackage[eulergreek]{sansmath}
\pgfplotsset{
  tick label style = {font=\sansmath\sffamily},
  every axis label = {font=\sansmath\sffamily},
  legend style = {font=\sansmath\sffamily},
  label style = {font=\sansmath\sffamily},
}
\pgfplotsset{compat=1.9}
\usepackage[separate-uncertainty=true,product-units=power]{siunitx}
\usepackage[caption=false]{subfig}
\usepackage{tablefootnote}

\usepackage{xcolor}
\definecolor{lightBlue}{rgb}{0.52941,0.58431,1.00000}%
\definecolor{lightRed}{rgb}{1.00000,0.52941,0.52941}%

\begin{document}

\preprint{APS/123-QED}

\title{Near-Surface Electrical Characterisation of Silicon Electronic Devices Using Focused keV Ions}
% Force line breaks with \\
\author{S. G. Robson}
\email{simonr1@student.unimelb.edu.au}
\affiliation{ 
Centre for Quantum Computation and Communication Technology, School of Physics, University of Melbourne, Parkville VIC 3010, Australia
}

\author{P. R{\"a}cke}%
\affiliation{ 
Leibniz Institute of Surface Engineering (IOM), Permoserstr. 15, D-04318 Leipzig, Germany
}%
\affiliation{
Leibniz Joint Lab `Single Ion Implantation', Permoserstr. 15, D-04318 Leipzig, Germany
}

\author{A. M. Jakob}
\author{N. Collins}
\affiliation{ 
Centre for Quantum Computation and Communication Technology, School of Physics, University of Melbourne, Parkville VIC 3010, Australia
}

\author{H. R. Firgau}
\author{V. Schmitt}
\author{V. Mourik}
\author{A. Morello}
\affiliation{ 
Centre for Quantum Computation and Communication Technology, School of Electrical Engineering and Telecommunications, UNSW Sydney, Sydney NSW 2052, Australia
}

\author{E. Mayes}
\affiliation{
RMIT Microscopy and Microanalysis Facility, RMIT University, Melbourne VIC 3001, Australia
}

\author{D. Spemann}
\affiliation{ 
Leibniz Institute of Surface Engineering (IOM), Permoserstr. 15, D-04318 Leipzig, Germany
}%
\affiliation{
Leibniz Joint Lab `Single Ion Implantation', Permoserstr. 15, D-04318 Leipzig, Germany
}

\author{D. N. Jamieson}
\affiliation{ 
Centre for Quantum Computation and Communication Technology, School of Physics, University of Melbourne, Parkville VIC 3010, Australia
}

\date{\today}% It is always \today, today,

\begin{abstract}
The demonstration of universal quantum logic operations near the fault-tolerance threshold establishes ion-implanted near-surface donor atoms as a plausible platform for scalable quantum computing in silicon. The next technological step forward requires a deterministic fabrication method to create large-scale arrays of donors, featuring few hundred nanometre inter-donor spacing. Here, we explore the feasibility of this approach by implanting low-energy ions into silicon devices featuring an enlarged \SI{60x60}{\micro\metre} sensitive area and an ultra-thin \SI{3.2}{nm} gate oxide  -- capable of hosting large-scale donor arrays. By combining a focused ion beam system incorporating an electron-beam-ion-source with in-vacuum ultra-low noise ion detection electronics, we first demonstrate a versatile method to spatially map the device response characteristics to shallowly implanted \SI{12}{keV} $^1$H$_2^+$ ions. Despite the weak internal electric field, near-unity charge collection efficiency is obtained from the entire sensitive area. This can be explained by the critical role that the high-quality thermal gate oxide plays in the ion detection response, allowing an initial rapid diffusion of ion induced charge away from the implant site. Next, we adapt our approach to perform deterministic implantation of a few thousand \SI{24}{keV} $^{40}$Ar$^{2+}$ ions into a predefined micro-volume, without any additional collimation. Despite the reduced ionisation from the heavier ion species, a fluence-independent detection confidence of $\geq$99.99\% was obtained. Our system thus represents not only a new method for mapping the near-surface electrical landscape of electronic devices, but also an attractive framework towards mask-free prototyping of large-scale donor arrays in silicon.
\end{abstract}
\maketitle

\section{Introduction}
Devices based on complementary metal–oxide semiconductor (CMOS) technology, both in established applications for classical information technology and in emerging applications for quantum computing, depend on the precise control of electric fields in the near-surface region. In classical silicon transistors, the source-drain current flowing beneath a thin (\SI{<10}{nm}) gate oxide is modulated by the electric field from a surface gate \cite{Sze2012c}. The current flows in thin sheets, with the highest density closest to the Si-SiO$_2$ interface \cite{Rau1999}. With the transistor size inside modern processors now approaching \SI{5}{nm} \cite{IEEE_roadmap}, the concentration of fabrication defects on the nanoscale has become a critical issue. For example, local oxide thickness fluctuations, trapped interface charge, and atomic-scale interface roughness can strongly affect the device internal electric field, leading to device reliability issues and breakdown \cite{Asenov2002,Asenov2003}. In devices engineered for quantum applications, the electronic and nuclear spin of near-surface (\SI{<20}{nm} deep) donors in silicon can be used to encode quantum information \cite{kane}. Each atom's quantum state is programmed and read-out by electric fields from surface gates isolated from the substrate by a thin gate oxide \cite{Tosi2017}. Small-scale prototype devices have so far demonstrated nuclear coherence times exceeding \SI{30}{s} \cite{Muhonen2014}, gate fidelities beyond 99.9\% \cite{Dehollain2016}, and more recently, universal quantum logic operations in a three-qubit processor with >99\% fidelity \cite{Madzik2022}. To satisfy surface code error correction specifications, operational quantum devices will require a large number (>$10^6$) of donors arranged in a near-surface array \cite{Tosi2017}. Besides being an enormous fabrication challenge in its own right, careful control of the internal electric field profile is critical for these devices to perform as intended.

Among the possible fabrication pathways to incorporate near-surface donors in silicon, scanning probe lithography can be used to produce few-donor clusters \cite{Ivie2021} with sub-nm placement precision \cite{Fricke2021}. An alternative placement approach involves using low-energy ion implantation to produce well-separated individual donors. Already the industry standard for fabricating classical CMOS devices \cite{Poate2002a}, we recently extended this technique with the development of a deterministic ion implantation system that uses the internal electric field within a silicon device to register stochastic arrival-time single ion implantation events \cite{Jakob2021}. The system was used to demonstrate post-implant counted detection of a few thousand \SI{14}{keV} P$^+$ ions in Si with 99.85\% confidence, using an ion beam collimated to \SI{\sim 10}{\micro\metre} diameter by an in-situ microstencil \cite{Jakob2021}. The inherent uncertainty of each ion's final resting position in the silicon lattice is furthermore fully compatible with the constraints of the ``flip-flop" qubit architecture \cite{Tosi2017}, where $\SIrange{100}{500}{nm}$ inter-donor spacing is permitted. 

In this work, we investigate the possibility of enhancing this fabrication method to produce large-area arrays of single donors spanning tens of \si{\micro\metre} -- a key prerequisite for a large-scale silicon quantum processor. We employ focused ion beam (FIB), a well-understood analysis and fabrication technique that has been used within materials science for decades \cite{Orloff1993}. Typically incorporating a Ga liquid-metal ion source with a \SI{\sim 5}{nm} beam spot size, FIB is frequently used for milling microstructures in a variety of materials \cite{Li2021}. More recently, the use of light ion sources such as He and Ne has reduced the beam spot size to \SI{<0.5}{nm} \cite{Klingner2020}, leading to applications in photonics \cite{Manoccio2021} and quantum materials \cite{Moll2018}. Using a new generation of single ion detectors that feature an ultra-thin passivation gate oxide plus an enlarged sensitive area suitable for spin readout and control of donor arrays \cite{Tosi2017,Jakob2021}, we first scan a focussed probe of \SI{12}{keV} $^1$H$_2^+$ ions across the surface and measure the device's response as a function of beam position. This technique is commonplace for electrical device characterisation at few-\si{\micro\metre} depths using high-energy MeV ions \cite{Vittone2013}, but here its first use in the low keV regime is demonstrated. As such, this provides a unique method to evaluate the near-surface (\SI{<100}{nm}) electronic properties. For devices that meet appropriate quality acceptance criteria, we then reconfigure the system to perform counted \SI{24}{keV} $^{40}$Ar$^{2+}$ implants into the centre of the sensitive region, without the use of an additional mask, and calculate the ion detection confidence. This allows us to assess the future suitability of using the system as a potential new method to construct donor arrays in silicon.

\begin{figure*}
    \centering
    \includegraphics{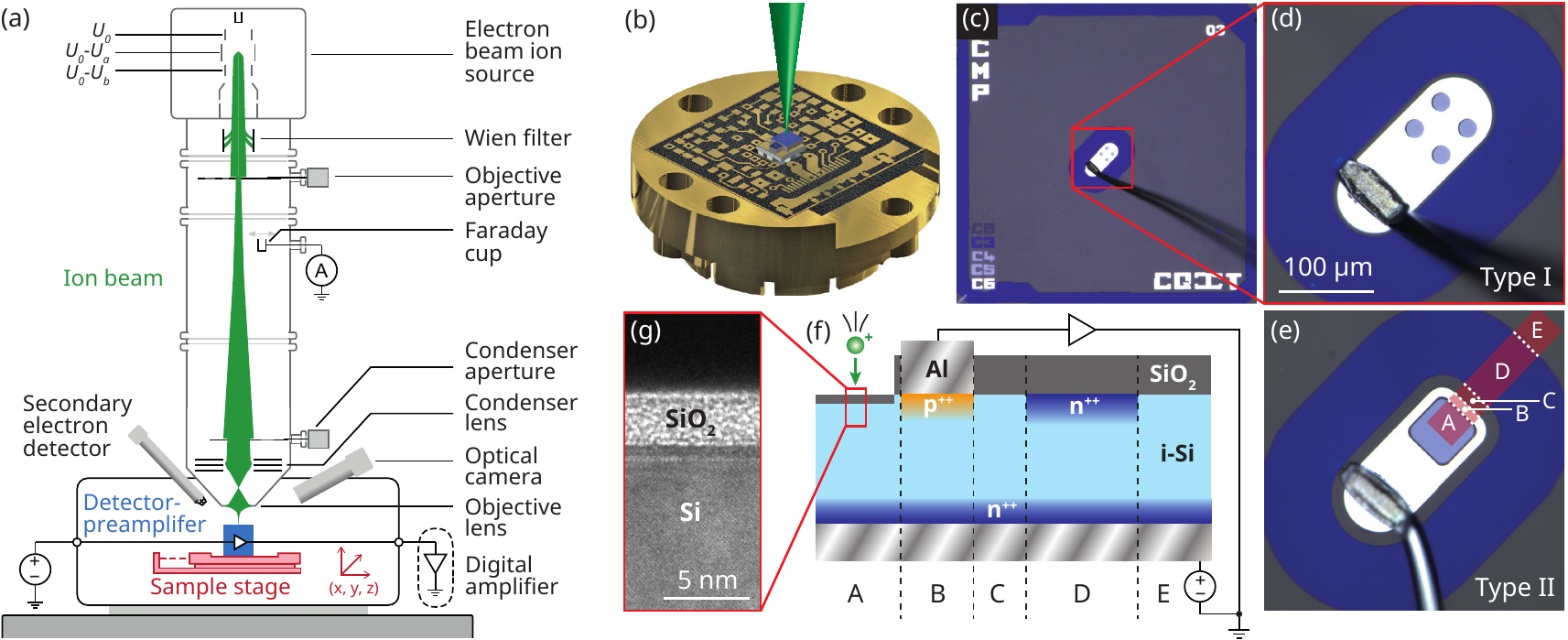}
    \caption{(a) Schematic of the FIB machine, equipped with an electron-beam-ion-source and integrated single ion detector electronics mounted on a precision sample stage. The ion beam profile is based on an actual ray tracing simulation \cite{Racke2020}. (b) Computer-aided design image of the single ion detector setup. It consists of a preamplifier circuit board housed in a metal case of 4 cm diameter. The ion beam (green) is focused onto an on-chip silicon p--i--n diode detector that is directly connected to a low noise junction field effect transistor (JFET). Both the detector and the JFET are mounted on a thermoelectric cooling element at the centre of the board. (c) Optical micrograph showing the top-view layout of the ion detector diode. Two different layouts of (d) $4\times$\SI{20}{\micro\metre}-diameter construction sites (CS) or (e) one central \SI{60x60}{\micro\metre} CS are available. (f) Cross-sectional schematic of the detector (as indicated in (e)) showing the sandwich-type p--i--n diode structure. The gate oxide inside the CS is between \SIrange{3}{6}{nm} thick, whilst the surrounding field oxide exhibits \SI{\sim 65}{nm} thickness. The aluminium electrodes are approximately \SI{200}{nm} thick. (g) Representative cross-sectional transmission electron microscope image of the CS of a detector. \label{fig:setup}}
\end{figure*}
\section{Methods}
\subsection{The Modified FIB System}
Experiments took place using a commercially available FIB machine (Raith GmbH) that has been modified for this work (schematic shown in Fig \ref{fig:setup}(a)). Full details of the machine are available in \cite{Racke2020}. Briefly, the conventional liquid metal ion source was replaced with an electron-beam-ion-source (EBIS) from DREEBIT GmbH \cite{Schmidt2009}. This is a versatile high-brightness gas plasma source that can produce ions in multiple charge states $q=\{1,2,...\}$ up to complete ionisation \cite{Ullmann2007}. In this study, Ar$^{q+}$ and H$_2^+$ molecule ions were employed. The base pressure in the ion source chamber was kept below \SI{e-9}{mbar}. The source potential can vary between \SIrange{3}{20}{kV}, but was generally set to \SI{12}{kV} for the these experiments, resulting in a kinetic ion implantation energy of $E = q\times 12 \hspace{2pt}\si{keV}$. Mass and charge-state selection occur via an integrated Wien filter, with a retractable Faraday cup located at the source output for ion beam diagnostics. Ion focusing optics comprise of objective and condenser apertures ranging from \SIrange{1}{200}{\micro\metre} in size, selectable via an electrically-driven aperture stage. An octupole stigmator ensures a circular ion beam profile. The beam is de-magnified by the FIB lens system, and can be scanned across the sample stage using the integrated high-speed pattern generator. The working distance is \SI{10}{mm}. Additionally, the interferometrically controlled sample stage can be positioned to \SI{<2}{\nano\metre} accuracy across its entire \SI{200}{mm} travel range and provides an alternative precision sample stepping method. Coarse lateral ion beam alignment (\SI{\sim 10}{\micro\metre}) is accomplished by an optical camera, and secondary electron imaging via an integrated Everhart-Thornley detector enables precision sample alignment to within \SI{\sim 20}{\nano\metre}. Beam currents as low as \SIrange{0.1}{100}{ion/s}, compatible with deterministic ion implantation, are achieved by combining direct filament current control and Wien filter detuning. 

\subsection{Ion Detection}
The ion beam is focused onto a specially-configured silicon diode (referred hereafter as a detector) mounted on a miniature preamplifer printed circuit board. The assembly is housed within a metal case (see Fig. \ref{fig:setup}(b)) which is fixed to the FIB sample stage. The detector design incorporates a localised p-doped top electrode region and a uniform n-doped back contact in \SI{520}{\micro\metre}-thick uniform high purity [001] float zone Si (n-type, \SI{9250}{$\Omega$ cm}) to form a vertical ``sandwich type" p--i--n structure, as shown in Fig. \ref{fig:setup}(c)-(f). The electrodes are metallised with \SI{200}{nm}-thick Al to form Ohmic contacts for wire bonding. Ions are designed to be implanted into the centre of the detector (region A, referred hereafter as the "construction site", CS), which is covered by a thermally grown passivation gate oxide of between \SIrange{3}{6}{nm} nominal thickness. The remainder of the detector surface is covered by a thermally grown field oxide of \SI{\sim65}{nm} thickness (regions C, D, E in Fig. \ref{fig:setup}(f)). The top electrode (region B) is surrounded by a buried floating n-doped guard ring (region D) to screen against excess leakage current from surface, interface, and bulk defect states from the surrounding region \cite{Evensen1993}. Two detector models are studied in this work: a previous generation incorporating four circular CS, each with \SI{20}{\micro\metre} diameter (``Type I", Fig. \ref{fig:setup}(d)) and a \SI{5.9}{nm}-thick gate oxide; and a new generation utilising a single CS of \SI{60}{\micro\metre} edge length (``Type II", Fig. \ref{fig:setup}(e)) with either a \SI{5.2}{nm} or \SI{3.2}{nm}-thick gate oxide. Full details on the general detector design and fabrication process are described elsewhere \cite{Jakob2021}. Key properties of the specific detectors employed in this work are summarised in Table \ref{tab:detectors}.

Each ion impact event is measured using the ion beam induced charge (IBIC) technique \cite{Breese2007}. As each ion strikes the detector, a cascade of electron-hole (e-h) pairs is created along its deceleration trajectory. The number of e-h pairs produced in the sensitive silicon detection volume $n_0$ is proportional to the remaining kinetic ion energy after passing through the oxide passivation ``dead" layer $E'=E-\delta E$ and is made up of contributions from the primary ion as well as those of respective recoiled target atoms. Through a process of diffusion and drift (from the internal electric field or with an additional external reverse bias), the e-h pairs become separated. A fraction may recombine at trapping sites in the silicon bulk as well as at the Si-SiO$_2$ interface, resulting in $n < n_0$ e-h pairs reaching the electrodes. By connecting the detector to a charge-sensitive preamplifier circuit, the electrical impulse due to ion-induced charge movement towards the electrodes can be measured. The detector's ion detection ability is typically quantified by the charge collection efficiency $\eta=n/n_0$.

In this work, an ultra-low noise preamplifier based on the design of Bertuccio et al. \cite{Bertuccio1994} is employed. It incorporates a forward-biased junction field effect transistor (JFET) \cite{MoxtekInc.2004} which, together with the detector, is mounted on an integrated thermoelectric cooling element (see Fig. \ref{fig:setup}(b)). The detector is operated under reverse bias by applying +\SI{10}{V} to the lower electrode and connecting the top electrode to the JFET gate via a wire bond. Further signal processing occurs entirely within the integrated preamplifier. Modest cooling to \SI{-10}{\celsius} is applied during operation, resulting in a typical root-mean-square (r.m.s.) noise of \SI{\sim 70}{eV} \cite{Jakob2021}. The pre-amplified signal output is then fed into an Amptek PX5 digital pulse processor \cite{AmptekInc.2017} which performs trapezoidal pulse shaping ($\tau=\SI{9.6}{\micro\second}$ peaking time), amplification, and digitisation. Channel-to-energy conversion is accomplished by comparing the ion response against that of a $^{57}$Co calibration radionuclide that emits characteristic Fe \SI{6.40}{keV} K$_\alpha$ and \SI{7.06}{keV} K$_\beta$ X-ray photons. The correlation between the known ion beam position (provided directly from the FIB pattern generator) and the detected signal from each ion impact is realised using a custom-written NI LabVIEW program.

\begin{table}[]
\caption{Summary of the individual properties of each detector used in this work\label{tab:detectors}}
\begin{minipage}{\columnwidth}
\begin{tabular*}{\columnwidth}{c @{\extracolsep{\fill}} ccccc}
Detector & Wafer ID & Gate oxide & Field oxide & CS type \\
name           &           & thickness (nm)            & thickness (nm)        & \\\hline
A-1         & A        & 5.2\footnote{An additional 500 nm thick layer of SU-8 resist was deposited on the gate oxide and formed into a calibration pattern via electron beam lithography.}                     & 62                         & II             \\
B-1         & B       & 5.9                       & 65                         & I           \\
C-1         & C       & 3.2                       & 68                         & II             \\
C-2         & C        & 3.2                       & 68                         & II             \\

\end{tabular*}
\end{minipage}
\end{table}

\section{Results and Discussion}
\subsection{Setup Characterisation}
\subsubsection{FIB Performance}

\begin{figure}[h!]
	\centering
    \resizebox{\columnwidth}{!}{%
    \includegraphics{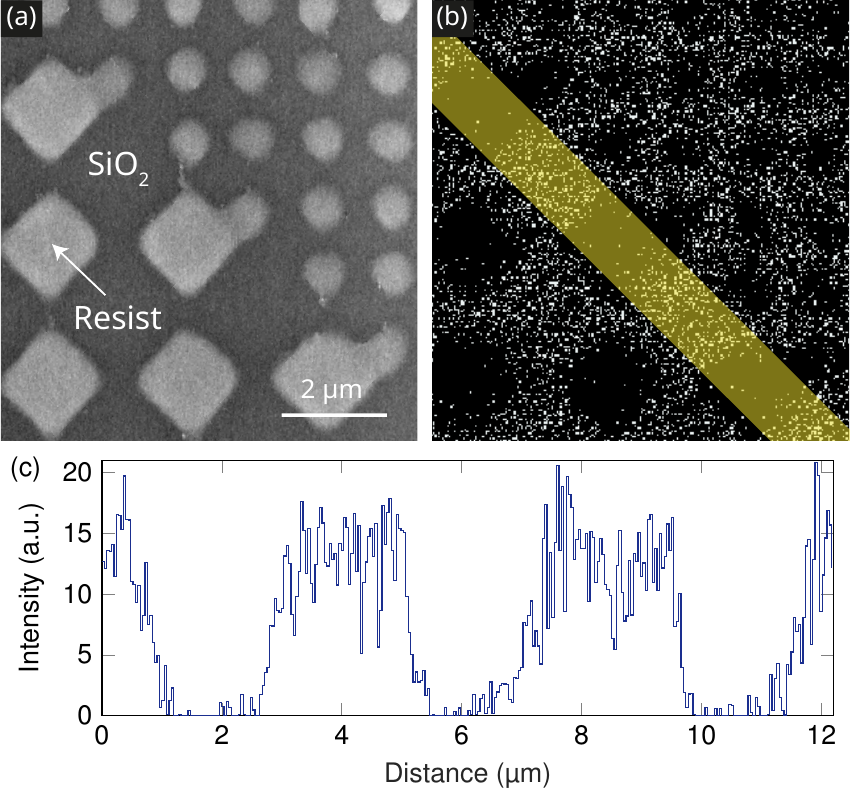}
}
\caption{(a) Secondary electron image taken inside the construction site of detector A-1, showing the checkerboard pattern formed by a 500 nm thick SU-8 resist layer deposited on the gate oxide. Squares of \SI{2}{\micro\metre} and \SI{1}{\micro\metre} pitch can be seen. (b) Corresponding $256\times256$ pixel spatially-resolved single ion impact event map, created by a focused beam of \SI{6}{keV} H$_2^+$ ions. A total of 6521 ion impacts were recorded, with each white pixel corresponding to a single detected ion impact event. The ion impact rate was approximately \SI{20}{ion/s} and the dwell time was fixed at \SI{8}{ms}. Objective and condenser aperture sizes were set to \SI{1}{\micro\metre} and \SI{10}{\micro\metre}, respectively. (c) Linescan extracted from the yellow-shaded line in (b) and integrated over its width. \label{fig:calibration}}
\end{figure}

Evaluation of the ion beam profile took place with detector A-1, where an additional \SI{500}{nm} of SU-8 resist was deposited on the surface and selectively developed into a calibration pattern using electron beam lithography. A checkerboard layout with pitch size ranging from \SIrange{4}{0.5}{\micro\metre} was formed inside the CS, as shown in Fig. \ref{fig:calibration}(a). The detector was subsequently placed inside the FIB machine, where a \SI{6}{keV} $^1$H$_2^+$ beam was then scanned over a \SI{10 x 10}{\micro\metre} area inside the CS. Light ions are employed for the initial characterisation because they produce less sample damage and greater ionisation than heavier species for the same kinetic energy \cite{Vittone2013}. Each $^1$H$_2^+$ molecule ion dissociates quasi-instantaneously upon surface impact, due to the energy transfer exceeding the binding energy by several orders of magnitude \cite{Cooks1990}. The constituent ions share the total kinetic energy in equal parts (\SI{3}{keV} per H$^+$) and decelerate simultaneously inside the detector. Ions that strike areas protected by the resist cannot produce signal pulses, as its thickness is large enough to prevent incident ions from entering the sensitive detector volume \cite{Ziegler2010}. In resist-free areas, the average penetration depth is \SI{\sim 47}{nm} below the Si-SiO$_2$ interface, and a combined sum of \SI{1560}{\text{e-h pairs}} is produced per H$_2^+$ ion \cite{Ziegler2010}. The shallow implantation depth compared to the overall detector thickness causes the IBIC signal pulse to be dominated by electron drift towards the back electrode, whereas holes contribute to only a minor degree \cite{Vittone2008}.

Fig. \ref{fig:calibration}(b) demonstrates a spatially-resolved map of ion-induced signal events captured with optimised FIB column parameters. A comparison against the actual calibration pattern visualised via scanning electron microscopy (Fig. \ref{fig:calibration}(a)) shows that the overall shape appears well-reproduced by the system. The majority of detected ion impact events originate from resist-free areas, and the pattern aspect ratio is preserved, indicating optimised ion beam stigmator settings. However, some residual scattered events are also seen to occur within resist-covered areas. To better understand this finding, intensity line profiles were extracted from the ion impact event map, with an example shown in Fig. \ref{fig:calibration}(c). The left-hand edge of each plateau (with respect to Fig. \ref{fig:calibration}(c)) appears to have a consistently greater spread than the right-hand edge, with this highly directional behaviour indicating unwanted residual FIB coma effects. This can be addressed by further fine adjustment of the ion optical alignment \cite{Racke2020}. The beam spot of the system was thus determined by fitting an error function to the unaffected right-hand edge. An average FWHM of \SI{181\pm14}{nm} was obtained along an extended edge feature. This is a substantial improvement over other EBIS-based FIB machines, where values of between \SIrange{500}{1000}{nm} have been reported \cite{Ullmann2007,Guillous2016,Schmidt2018EBIS-basedMicro-beams}. However, this does not represent the lower resolution limit of our system. An even smaller spot size can be obtained by further reducing the FIB aperture size to below \SI{100}{nm}, despite a predicted \SI{<1}{fA} beam current that would not provide sufficient statistics for conventional secondary electron-generated images. Custom-fabricated nanoapertures are currently under evaluation for this purpose \cite{Emmrich2016}.

\subsubsection{Ion Energy Measurement \& Mapping}
\begin{figure}
    \centering
    \resizebox{\columnwidth}{!}{%
    \includegraphics{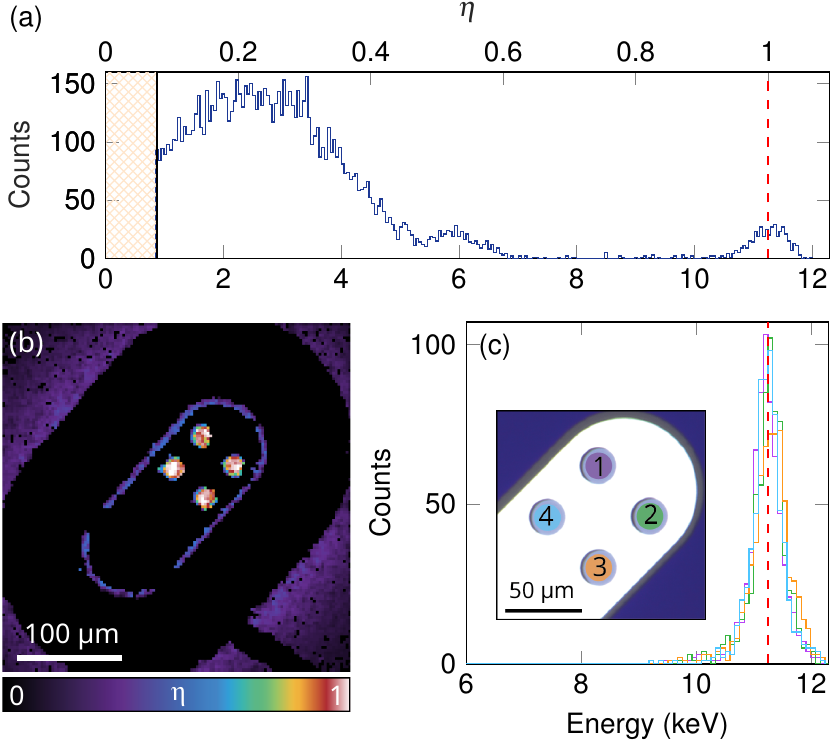}
    }
    \caption{(a) Signal pulse spectrum of detector B-1 exposed to a scanned 12 keV H$_2^+$ beam. The three peaks correspond to different regions of the device with varying $\eta$, as discussed in the main text. The hatched interval $[0,0.84]$ keV indicates the noise discriminator regime. The red dashed line indicates an average charge collection efficiency of $\langle\eta\rangle=1$. (b) Spatially resolved $128\times128$ pixel map of the data presented in (a). The dark strip extending to the bottom-right corner is due to the shadowing effect of the  \SI{20}{\micro\metre}-thick wire bond attached to the metal top electrode. (c) Signal pulse spectra extracted from detailed scans performed inside each construction site (as indicated in each highlighted area in the inset optical image). Each spectrum is comprised of $\sim$650 single ion impact events.}
    \label{fig:noise_floor}
\end{figure}

To confirm correct operation of the system, a \SI{12}{keV} $^1$H$_2^+$ ion beam was scanned over a detector with a previously-measured response to similar energy $^1$H$_2^+$ ions \cite{Jakob2021} (detector B-1, Type I) and the signal pulse amplitudes were recorded. In the absence of incident ions, a r.m.s. noise of \SI{\sim 110}{eV} was measured. This is about \SI{40}{eV} greater than other detectors from the same wafer \cite{Jakob2021} and can be attributed to an additional capacitance at the JFET input gate from a second detector that was connected in parallel during this experiment. However, this is sufficient for high confidence low-energy single ion detection, as discussed later. A total of $10,000$ single ion impact events were recorded, with the resulting signal pulse spectrum shown in Fig. \ref{fig:noise_floor}(a). The spectrum is dominated by a broad main peak centred at \SI{2.5}{keV}, with a fraction of its lower energy side cut off due to the noise discriminator threshold. On the high-energy side, a separate small signal peak occurs at \SI{5.8}{keV} and approximately \SI{700}{eV} FWHM. One further isolated small peak, centred at \SI{11.3}{keV} and approximately \SI{600}{eV} FWHM, can also be identified.

The corresponding spatially-resolved charge signal pulse map is shown in Fig. \ref{fig:noise_floor}(b) and reveals the origin of each peak. Each pixel is colour-coded according to the charge collection efficiency $\eta$ determined from the pulse recorded at the given location. The overall map features agree well with the optical top-view micrograph of the same device shown in Fig. \ref{fig:setup}(d). As expected, the isolated highest-energy signal peak originates inside the CS array (region A in Fig. \ref{fig:setup}(f)). Here, the initial ion energy loss from stopping in the thin gate oxide (see Table \ref{tab:detectors}) is nearly negligible. Along with the low bulk defect concentration in the high-purity silicon wafer and the high-quality thermal gate oxide interface \cite{Jakob2021,Johnson2010}, this maximises the number of e-h pairs produced in the sensitive volume, resulting in $\eta \approx 1$. For a \SI{12}{keV} H$_2^+$ molecule ion, $\sim$95\% of its kinetic energy is dissipated in electronic stopping by e-h pair generation according to the model of Funsten and Ziegler \cite{Funsten2001,Ziegler2010}. With the signal peak centre located at \SI{\sim 11.3}{keV}, our results are consistent with this model, and also with previous experimental results obtained with a collimated beam from a conventional plasma-filament ion source \cite{Jakob2021}, thus confirming the correct functionality of our system. Additional spectra obtained from independently scanning each individual CS are also shown in Fig. \ref{fig:noise_floor}(c). There appears to be minimal variation in $\eta$ between each CS, and the absence of isolated events outside the main signal peak indicates negligible artificial influences -- e.g. from ion scattering or gate oxide effects such as thickness fluctuations and surface debris. 

Outside of the CS, the top metal electrode pad (region B) yields no single ion detection events, agreeing with TRIM simulations that predict 100\% of \SI{12}{keV} H$_2^+$ molecule ions should stop completely within the metal layer \cite{Ziegler2010}. Next, the narrow undoped region between the top electrode and the n-guard ring (region C) gives rise to the satellite signal peak located at \SI{6}{keV}. Here, the ions experience increased stopping in the thicker field oxide (see Table \ref{tab:detectors}) and lose approximately half of their initial kinetic energy before reaching the sensitive detector volume \cite{Ziegler2010}. Hence, only approximately half the number of e-h pairs is generated in this region, compared to in region A. Further, no signal events are recorded within the n-guard ring area (region D) due to the very high phosphorous doping concentration (\SI{3e19}{cm^{-3}}), representing a volume of very effective e-h pair recombination. 

Finally, the outer area (region E) yields the dominant signal peak in Fig. \ref{fig:noise_floor}(a), with the peak size coming from its relatively large area fraction of the detector. Here, the n-guard ring strongly attenuates the internal reverse bias drift field, rendering the slower charge carrier diffusion process (minority carrier lifetimes typically \SI{\sim100}{ns}, compared to ps drift times \cite{Breese1993}) as the dominant transport mechanism. The spatially isotropic nature of e-h pair diffusion, combined with lifetime-limiting interface and bulk defects, results in a quick decline of $\eta$ with growing lateral distance from the n-guard edge. However, we emphasise that ions are not actually designed to be implanted and detected in this region. Instead, the earlier result showing $\eta \rightarrow 1$ within each CS (region A) is the key conclusion to draw from this initial experiment.

\subsection{Defect Measurement and Analysis}
\begin{figure*}
\centering
    \includegraphics{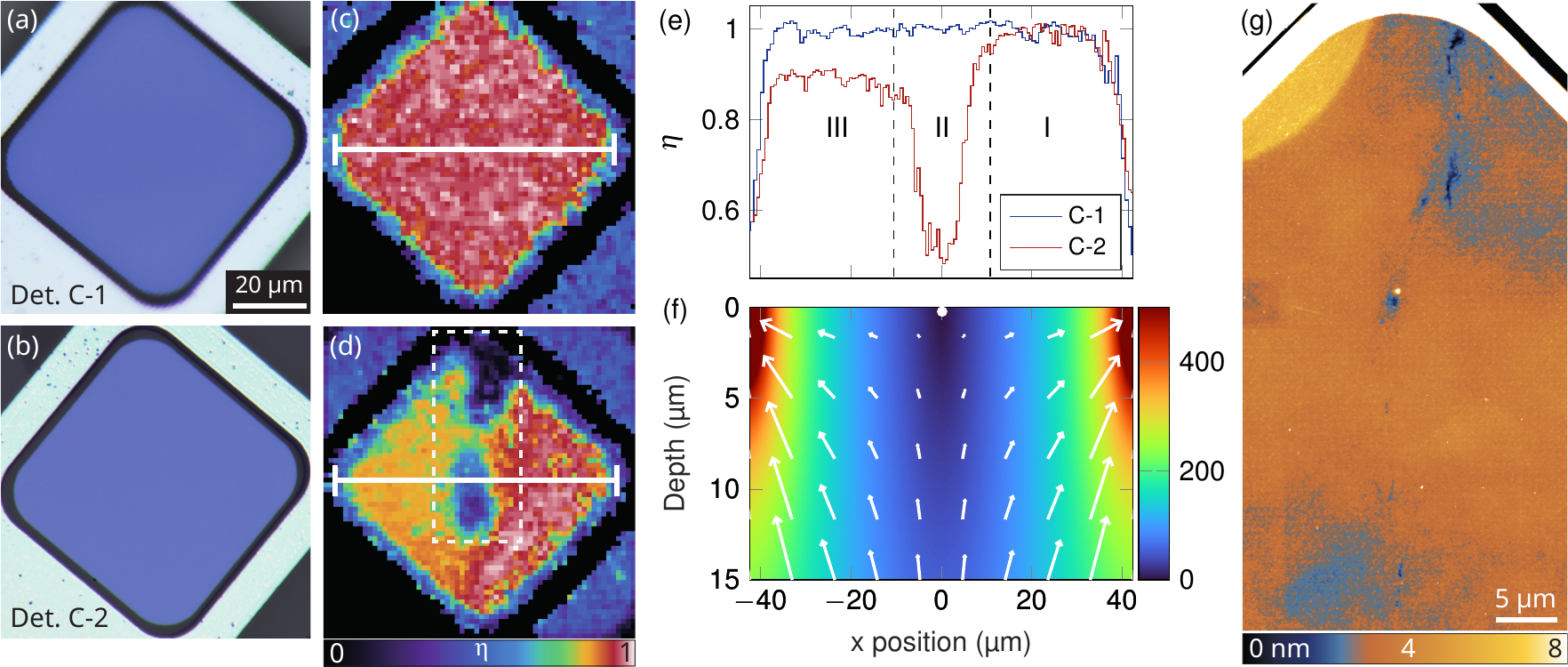}
\caption{$100\times$ optical micrographs showing the construction sites of (a) detector C-1 and (b) detector C-2. (c), (d) Corresponding spatially-resolved $256\times256$ pixel signal pulse maps captured using a scanned \SI{12}{keV} H$_2^+$ beam with a \SI{\sim 20}{ion/s} incidence rate and a fixed \SI{8}{ms} dwell time. A two-pixel binning algorithm was applied to compensate for some void pixels. (e) Line profiles of $\eta$, extracted along the diagonal of the construction site of each detector (white lines in (c) and (d)). Regions of similar $\eta$ for detector C-2 are indicated and discussed in the main text. (f) COMSOL\textsuperscript{\textregistered} simulation \cite{COMSOL} showing the strength and direction of the internal electric field (\si{V/cm}) inside an ideal detector, taken along the diagonal cross-section of the construction site. An external reverse bias of \SI{-10}{V} was applied. Peak doping concentrations of \SI{1e16}{cm^{-2}} were specified in the p and n ohmic contact regions against a \SI{1e11}{cm^{-2}} n-type background. An interface defect density of $D_{it}=\SI{1e11}{{cm}^{-2}{eV}^{-1}}$ and a fixed oxide charge of $Q_f=\SI{1e9}{cm^{-2}}$ were assumed in the gate oxide. The white dot at the image centre indicates the expected average implantation depth of \SI{12}{keV} H$_2^+$ ions according to TRIM calculations \cite{Ziegler2010}. (g) Topography map of detector C-2 obtained with atomic force microscopy. The scan area is indicated by the dashed white rectangle in (d). \label{fig:CS_CCE}}
\end{figure*}

Turning to our application to silicon donor-based quantum computing, we now evaluate Type II detectors incorporating a large-area CS needed to host large-scale donor arrays. We will test the key array formation requirement: all implanted ions must be able to be detected with a high degree of confidence at their predefined target locations. More precisely, $\eta$ must meet two criteria in the CS: (1) the average value must be near-unity, and (2) it must be spatially homogeneous. As previously introduced, the IBIC technique is inherently sensitive to defects located in the bulk \cite{Auden2015a,Vittone2016} and at interfaces \cite{Vizkelethy2006,Tonigan2021a} that act as trapping/recombination centres for ion-induced free e-h pairs. Especially for criterion (2), local effects such as gate oxide thickness fluctuations or residual nanoscale surface debris can contribute towards increased ion stopping before an ion reaches the sensitive detector volume, leading to highly localised areas of reduced $\eta$. This cannot be readily evaluated with broad-beam ion implantation, and instead a scanned focused ion microprobe must be used. Furthermore, low kinetic energy keV ions are needed (unlike in conventional MeV IBIC experiments \cite{Vittone2013}) to ensure that a similar depth scale to that used for eventual donor array ion implants is probed.

To evaluate the presence of such defects, two representative detectors from the same fabrication wafer (see Fig. \ref{fig:CS_CCE}(a) and (b)) were selected at random and mapped with a scanned \SI{12}{keV} H$^+_2$ beam. Fig. \ref{fig:CS_CCE}(c) shows a high-resolution map of $\eta$ in detector C-1. The CS exhibits ideal signal characteristics and clearly fulfills both criteria. A linescan taken diagonally along its cross-section (Fig. \ref{fig:CS_CCE}(e)) confirms fluctuations in $\eta$ of $<3$\%. These can be attributed to the statistical nature of a multi-atom collision cascade each ion experiences, due to (i) variations in the residual kinetic ion energy after transmitting the gate oxide, (ii) variations in the electronic ion stopping fraction in the sensitive region, and (iii) Fano statistics of e-h pair generation due to electron and hole scattering in the silicon lattice \cite{Steinbauer1994}. This is also the origin of the signal peak width visible in Fig. \ref{fig:noise_floor}(c). The sharp decrease in $\eta$ at the CS perimeter is due to a sudden transition between the gate oxide and the thicker surrounding field oxide.

The situation is markedly different for detector C-2. As shown in Fig. \ref{fig:CS_CCE}(d) and (e), three distinct regions of different $\eta$ are observable within the CS. The right-hand third (region I) exhibits comparably uniform signal characteristics with near-unity $\eta$, similar to detector C-1. However, the remaining regions indicate the presence of defects. Two pockets of degraded $\eta<0.5$ occur at the top and in the centre of the CS (region II). They span approximately $10\times20$ \si{\micro\metre} in lateral dimensions. In the left-hand third of the CS (region III), a more uniform spatial detector response is apparent, albeit with a reduced average $\eta\approx\SI{0.9}{}$. To better understand the charge collection dynamics, a COMSOL\textsuperscript{\textregistered} simulation \cite{COMSOL} of the electric field $\mathbf{E}$ inside an ideal ($\eta=1$) detector was run, with the results shown in Fig. \ref{fig:CS_CCE}(f). $\mathbf{E}$ is largely uniform and vertically aligned deep inside the detector, but increasingly re-aligns horizontally in the vicinity of the surface. This is accompanied by a lateral gradient in the absolute field strength, with a near field-free region existing in the centre of the CS to a depth of a few \si{\micro\metre}. This is many times the ion implantation depth (\SI{88\pm 30}{nm} \cite{Ziegler2010}), suggesting that in this region, the motion of ion-generated free e-h pairs is initially characterised by diffusion and only afterwards becomes dominated by drift transport in the deeper silicon bulk. Remarkably, a high-quality Si crystal with no obvious point/extended defects (see Fig. \ref{fig:setup}(g)) and a well-passivated surface from a high-quality thermal oxide \cite{Fleetwood1993} appears sufficient to create an environment with diffusion lengths spanning tens of \si{\micro\metre}, as exhibited by detector C-1. This is clearly not the case in detector C-2.

The shallow ion probing depth suggests that the physical origin of the defective regions in detector C-2 may be surface-related. Atomic force microscopy was thus performed in the affected area of the CS, as indicated in Fig. \ref{fig:CS_CCE}(d)). The corresponding topography map is shown in Fig. \ref{fig:CS_CCE}(g). A hairline crack extending from the top corner to the centre of the CS can be identified, with lateral dimensions varying between a few hundred nm to a few \si{\micro\metre}. In some areas, the crack has a depth of up to \SI{4}{nm}. This is comparable to the thermal gate oxide thickness, suggesting that an error during the fabrication process (e.g. residual surface debris present during spin coating or a mechanical scratch from wafer handling) resulted in its local etching. The complete removal of a thermally-grown oxide and subsequent exposure of the underlying silicon material leads to the formation of a low-quality native oxide (\SIrange{2}{3}{nm} thickness), typically characterised by defect interface trap densities $D_{\mathrm{it}}>\SI{e13}{{cm}^{-2}{eV}^{-1}}$ and fixed oxide charge densities $Q_{\mathrm{f}}>\SI{e12}{cm^{-2}}$ \cite{Hori1997a}. These values lie about two orders of magnitude higher than those measured in a typical thermal gate oxide \cite{Johnson2010,Pla2012}, thus representing a concentrated region of effective e-h pair trapping and recombination. The corresponding signal pulse map (Fig. \ref{fig:CS_CCE}(d)) clearly reflects a reduced charge carrier lifetime in and around this damaged region, where only \SIrange{20}{50}{\%} of the ion-induced e-h pairs are collected at the detector electrodes. 

The non-uniform ion detection response exhibited by detector C-2 therefore makes this device unsuited to its proposed application in large-scale donor array formation for silicon quantum computing, clearly demonstrating the crucial need for a low interface density passivation thermal oxide in next-generation silicon quantum computing devices. However, this appears to be an isolated case, with other detectors from the same wafer also tested (not shown) and confirmed to be fit for purpose. Moreover, these results demonstrate the utility that a scanned keV ion probe provides in understanding and troubleshooting the near-surface device behaviour. The versatility of this system should also be pointed out in allowing not only analysis of silicon-based electronic devices, but potentially those based on other IBIC-compatible platforms, such as SiC, diamond, and III-V materials \cite{Vittone2013}.

\begin{table*}[]
\caption{Properties of selected ion species (and in some cases, varied kinetic energies) when implanted through a \SI{3.2}{nm}-thick SiO$_2$ layer into (100) Si, as modelled by Crystal-TRIM \cite{Posselt1992}. The implantation depth is calculated from the Si-SiO$_2$ interface. The number of e-h pairs is calculated in the sensitive detection volume, with a \SI{640}{eV} noise floor assumed to determine the ion detection confidence.
\label{tab:quantum}}
\begin{tabular*}{\textwidth}{c @{\extracolsep{\fill}} ccccccc}
Species & $E_{\mathrm{kin}}$ (keV)  & Implantation depth $\pm$ straggle (nm) & Lateral straggle (nm)   & $n_0$ (e-h pairs) & $\Xi(q_{\mathrm{t0}})$  (\%) & $Y_\mathrm{dop}$ (\%)  \\\hline %\hline
$^1$H$_2^+$           & 12            & 86.9 $\pm$ 31.2   & 40.3          & 3112  & $\geq$99.99 &   99.91       \\
$^{40}$Ar$^{3+}$      & 36            & 40.3 $\pm$  17.6  & 14.4          & 2764  & $\geq$99.99 &   99.99     \\
$^{40}$Ar$^{2+}$      & 24            & 27.8 $\pm$ 13.0   & 10.6          & 1695  & $\geq$99.99 &   99.95     \\
$^{40}$Ar$^+$         & 12            & 14.6 $\pm$ 7.8    & 6.3           & 625   & 99.84  &   99.68  \\ %\hline
$^{31}$P$^+$          & 14            & 19.7 $\pm$ 10.7    & 9             & 1108   & 99.98 & 99.62         \\
%$^{31}$P$^+$       & 15              & 17.0 $\pm$ 9.5    & 8             & 736   & 99.89$^{*)}$  &     \\
$^{31}$P$^+$          & 9             & 12.9 $\pm$ 7.7    & 6.5             & 643   & 99.85 & 98.98       \\
%$^{209}$Bi$^+$      & 14              & 12.3 $\pm$ 2.7    & 2.9           & x   & x    &
%x%Sb      & 51     & 12        & 11.3 $\pm$ 3.7    & 3.4           & 579      & 97.16 &               \\
%Bi      & 83     & 12        & 11.2 $\pm$ 2.5    & 2.7           & 389   & 96.31    &
\end{tabular*}
\end{table*}

\subsection{Mask-Free Deterministic Ion Implantation}
We now turn our attention towards reconfiguring the setup to enable future controlled implantation of individual $^{31}$P$^+$ ions for use in spin-based quantum computing. In these experiments, \SI{24}{keV} $^{40}$Ar$^{2+}$ ions ($\text{acceleration potential} = \SI{12}{kV}$) were employed.  Each ion now penetrates only \SI{28 \pm 13}{nm} below the Si-SiO$_2$ interface and generates an average of $\sim$\SI{1700}{e-h pairs} in the active detection volume \cite{Ziegler2010}. Although the electronic and spin properties of $^{40}$Ar do not make it compatible with quantum information processing in silicon, its ion stopping characteristics are very similar to the $^{31}$P donor-qubit \cite{Hopf2008}, providing a good estimate of the ion detection signal response and thus allowing a robust assessment of further adapting the FIB setup for this ultimate purpose.

\begin{figure}
    \centering
	    \resizebox{\columnwidth}{!}{%
    	\includegraphics{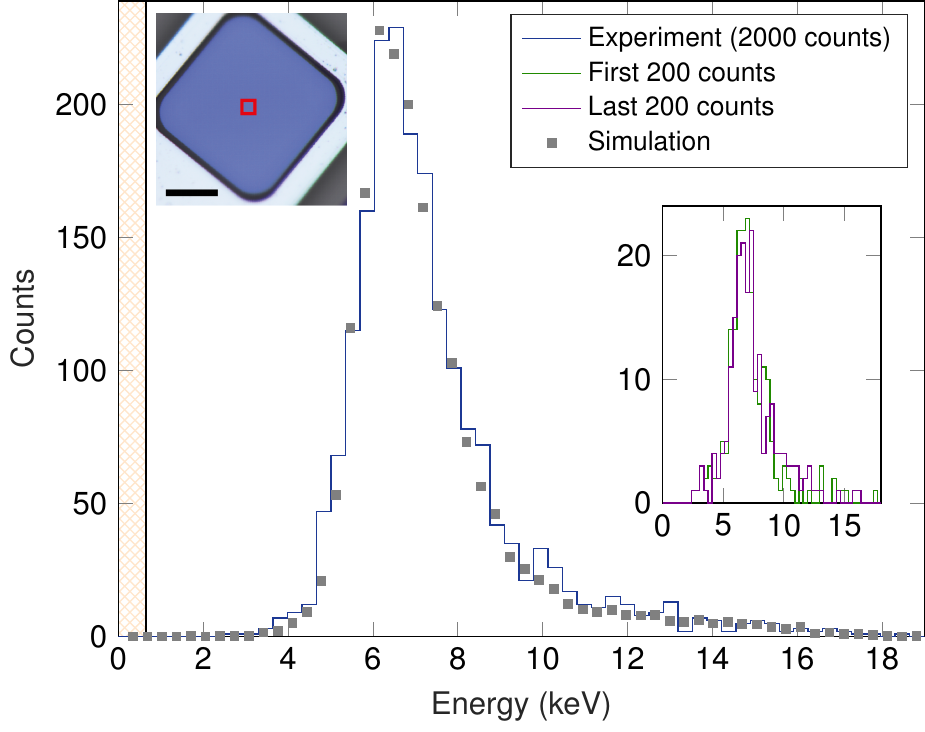}
    }
    \caption{Signal pulse spectrum of detector C-1 irradiated with a focused beam of $2,000$ \SI{24}{keV} Ar$^{2+}$ ions, raster-scanned over a $5\times5$ \SI{}{\micro\metre^2} area inside the construction site (as indicated in the left inset; $\text{scale bar} =  \SI{20}{\micro\metre}$). The dwell time was set to \SI{1}{ms} and the incidence rate was \SI{\sim 10}{ion/s}. The low level discriminator was set to \SI{640}{eV}, corresponding to the hatched region. Also shown are the results of normalised Crystal-TRIM simulations \cite{Posselt1992} of the same experiment, performed with $20,000$ ions. Right inset: sub-spectra from the first and last $200$ recorded counts.}
    \label{fig:argon}
\end{figure}

In a large-scale quantum computing device incorporating surface control gates \cite{Tosi2017,hill}, it is important to minimise the depth placement uncertainty (longitudinal straggle) of each donor. Too deep placement results in poor spin-dependent tunnel coupling of the donor electron, whereas too shallow placement causes the donor to be located adjacent to or in the oxide and consequently inactivated. The straggle can be reduced by lowering the implantation energy (see Table \ref{tab:quantum}). However, this also reduces the number of ion-induced e-h pairs and thus causes the amplitude of each charge pulse signal to lie near the noise threshold, thereby increasing the likelihood of $>1$ donor per implant site. The net result is a reduced ensemble coherence time from uncontrolled nearby donor-donor coupling. The solution is to reach an optimum balance between the placement uncertainty and the ion detection confidence $\Xi(q_{\mathrm{t0}})$, defined as the fraction of detected single ion events (counts above the noise discriminator threshold $q_{\mathrm{t0}}$) to the total number of incident ions \cite{Jakob2021}. To predict $\Xi(q_{\mathrm{t0}})$, a \SI{5 x 5}{\micro\metre} area located at the centre of the CS (left inset of Fig. \ref{fig:argon}) in detector C-1 was continuously scanned with a \SI{24}{keV} Ar$^{2+}$ beam until 2000 signal events were recorded. The resulting signal pulse spectrum is shown in Fig. \ref{fig:argon}. Compared to $^1$H$^+$, the heavier mass of $^{40}$Ar$^{2+}$ results in only \SI{25}{\%} of the initial ion kinetic energy being dissipated into electronic stopping \cite{Funsten2001}, shifting the peak centroid down to \SI{\sim 6.1}{keV}. The peak also appears broader due to a greater statistical variation in the fraction of electronic energy loss of each ion. Nevertheless, a clear interval between the lowest detected count (at \SI{2.5}{keV}) and the noise discrimination threshold (at \SI{0.64}{keV}) lacking any signal events is seen, strongly suggesting that 100\% of the incident ions were successfully detected. A robust assessment of $\Xi(q_{\mathrm{t0}})$ is done by comparing the experimentally obtained signal pulse spectrum with that generated by a Crystal-TRIM simulation \cite{Posselt1992} for an ideal ($\eta = 1$) silicon detector with the same gate oxide thickness. Good agreement between the experimental and simulated spectra is observed, allowing us to extract a detection confidence of $\Xi(q_{\mathrm{t0}})\geq99.99\%$. The effect of the accumulated ion fluence on the signal pulse spectrum \cite{Simon2007} was additionally considered by comparing the first and last 200 detected events, as shown in the right inset of Fig. \ref{fig:argon}. Within the statistical uncertainty of the two sub-spectra, no evidence of a systematic peak shift or broadening is seen, suggesting that even moderate ion-induced crystal damage (average Ar$^{2+}$ inter-ion spacing = \SI{112}{nm}) does not lead to a significant degradation in $\eta$. 

Compared to our previous detector generations incorporating a standard \SIrange{6}{8}{nm} gate oxide \cite{Jakob2021}, detector C-1 features an ultra-thin \SI{3.2}{nm} gate oxide (see Fig. \ref{fig:setup}(g)). Similar CMOS devices incorporating sub-\SI{5}{nm} gate oxides have been shown to be susceptible to elevated defect concentrations \cite{Stathis1998}. However, the near-unity $\eta$ value obtained here suggests that the density of interface defect traps and fixed oxide charge within this ultra-thin gate oxide is in fact comparable to that measured in previous-generation devices \cite{Madzik2022,Muhonen2014}. A thin gate oxide also provides an improved controlled silicon doping yield $Y_\mathrm{dop}$, defined as the fraction of detectable single ion impacts where the primary dopant ion also entirely transmits the gate oxide to stop inside the silicon crystal. Events where the primary ion stops inside the oxide may still produce a detectable signal, because forward-recoiled Si and O atoms from the oxide can also contribute to an overall charge signal pulse event. These events appear to be extremely rare for this detector generation, with Crystal-TRIM simulations \cite{Posselt1992} predicting $Y_\text{dop} = 99.95\%$ for \SI{24}{keV} Ar$^{2+}$ ions. In a previous detector generation, we reported a single ion detection confidence of $\Xi(q_{\mathrm{t0}})=99.85\%$ for \SI{14}{keV} $^{31}$P$^+$ ions, but the actual doping yield was limited to $Y_\mathrm{dop}=98.1\%$, due to $1.9\%$ of the incident $^{31}$P$^+$ ions stopping inside the \SI{6}{nm} gate oxide \cite{Jakob2021}. Based on the current results obtained with argon ions for this new detector generation, we expect a factor 5 enhancement to $Y_\mathrm{dop}=\SI{99.62}{\%}$ for \SI{14}{keV} $^{31}$P$^+$ ions, comfortably exceeding the latest estimates of tolerable donor qubit loss fault thresholds ($\SIrange{90}{95}{\%}$) \cite{Nagayama2017}. The installation of a $^{31}$P$^{+}$ ion source is forthcoming.

The lateral placement accuracy of each implanted ion is the additional component in determining our setup's suitability for scalable donor array fabrication in silicon, and can be split into two independent scattering processes. First, the uncertainty of each ion's lateral impact point on the detector surface comes from the fixed spot size of the FIB. Second, lateral ion straggling takes place inside the target material (see Table \ref{tab:quantum}) as a direct result of the random collision cascade each ion makes with target atoms upon stopping. Although the present \SI{181}{nm} (FWHM) spot size is not compatible with the sub-\SI{50}{nm} lateral donor placement constraints of the flip-flop qubit architecture \cite{Tosi2017}, we expect a further spot size reduction to \SI{<30}{nm} to be realistically achievable by employing custom-fabricated \SI{100}{nm} diameter lens apertures \cite{Emmrich2016}, as mentioned earlier. In this regime, effects from the spot size and the lateral ion straggling in silicon should contribute more or less equally to the overall ion placement precision. Furthermore, the expected reduction in beam current from the smaller aperture size can be offset by the highly adjustable nature of the EBIS emission intensity, yielding a similar expected on-target implantation rate to that used in the present experiments. Precision localisation of single ions has already been demonstrated with a masked approach, where a nanoaperture was milled in the tip of an atomic force microscope cantilever and placed directly above the sample to collimate the beam \cite{Pezzagna2010}. However, adapting this technique below the \SI{30}{nm} level may be technically challenging due to expected lateral and axial ion straggling effects from the aperture. Particularly for high aspect ratio nanoapertures, an unacceptably large proportion of ions can be implanted hundreds of nm away from their intended target locations \cite{Raatz2019}. This effect can also be aggravated by the accumulated ion fluence seen by the cantilever, whereby the nanoaperture can shrink and eventually completely close over time \cite{Pezzagna2010,Li2001a}. The use of a mask-less, focused implantation method, such as the FIB system employed in this work, may present a viable avenue around this.

Our system also has the advantage that highly charged ions (up to completely ionised) can readily be supplied by the EBIS. Especially for heavy donors such as $^{123}$Sb and $^{209}$Bi, the additional potential energy deposited by these ions in the detector \cite{Schenkel1999} can be utilised as an additional means to artificially increase $\Xi(q_{\mathrm{t0}})$. Alongside encouraging recent studies from systems that detect single ions just prior to the implantation event \cite{Racke2018,Lopez2019}, our results represent a promising alternative avenue towards the scalable engineering of near-surface donor qubit arrays with nanoscale placement precision.

\section{Conclusion}
We have introduced a FIB system equipped with an electron-beam-ion-source as well as ultra-low noise single ion detection technology to enable deterministic control over the number and position of implanted ions. The highly adjustable source can produce ions with energies between \SIrange{3}{300}{keV}, implanted at rates between \SIrange{0.1}{100}{ion/s}, focused down to a spot size of \SI{180}{nm} (FWHM). Through the ion beam induced charge principle, we use silicon p--i--n diode detectors to measure electron-hole pairs generated by each ion impact event and correlate this with the beam position. Using a new generation of ion detectors featuring an enlarged \SI{60x60}{\micro\metre} sensitive area and an ultra-thin \SI{3.2}{nm} gate oxide, we first spatially map the internal electric field to within \SI{100}{nm} of the surface using a rastered \SI{12}{keV} $^1$H$_2^+$ ion probe. The results highlight the destructive effect that localised regions of defect-rich native oxide play in suppressing the detection response, whereby >50\% of the initially diffused charge can be lost to associated interface states. Nevertheless, near-unity charge collection efficiency is measured in the sensitive area of functional detectors, thus confirming our devices are fit for purpose. Next, we modify the system to perform counted implantation of 2000 $^{40}$Ar$^{2+}$ ions at \SI{24}{keV} into a pre-defined $5\times5$ \SI{}{\micro\metre^2} area, obtaining $\geq$99.99\% single ion detection confidence. This approach foregoes the need for an mechanical implant mask that may introduce additional ion scattering. With an upcoming system upgrade to incorporate a $^{31}$P$^+$ ion source and yield a sub-\SI{30}{nm} FIB spot size, we aim to establish a viable method to fabricate large-scale near-surface $^{31}$P donor arrays in silicon for multi-qubit entanglement studies.

\section{Acknowledgements}
This work was funded by the Australian Research Council Centre of Excellence for Quantum Computation and Communication Technology (Grant No. CE170100012) and the US Army Research Office (Contract No. W911NF-17-1-0200). This work was performed in part in the NSW Node of the Australian National Fabrication Facility at UNSW Sydney. S.G.R, P.R, and A.M.J  acknowledge an Australia–Germany Joint Research Cooperation Scheme (UA-DAAD) travel scholarship that supported collaboration between partner institutions. S.G.R. acknowledges additional travel support from the Laby Foundation Pty Ltd. S.G.R. and H.R.F. acknowledge support from an Australian Government Research Training Program Scholarship. P.R. and D.S. gratefully acknowledge funding by the Leibniz Association (SAW-2015-IOM-1) and the European Union together with the S\"{a}chsisches Ministerium für Wissenschaft und Kunst (Project No. 100308873). Code for the Crystal-TRIM simulations performed in this work was developed by M. Posselt of Helmholtz-Zentrum Dresden-Rossendorf. The authors gratefully acknowledge the assistance of J. Bauer for performing electron microscopy, and A. Finzel (both IOM Leipzig) for performing atomic force microscopy. The views and conclusions contained in this document are those of the authors and should not be interpreted as representing the official policies, either expressed or implied, of the ARO or the US Government. The US Government is authorized to reproduce and distribute reprints for government purposes notwithstanding any copyright notation herein.

\newpage
\bibliography{references}% Produces the bibliography via BibTeX.

\end{document}